\documentclass[twocolumn,showpacs,preprintnumbers,amsmath,amssymb]{revtex4}

\usepackage{graphicx}
\usepackage{dcolumn}
\usepackage{bm}

\begin{document}


\title{Andreev Reflection and Pair Breaking Effects at the
Superconductor/Magnetic Semiconductor Interface}

\author{R.P. Panguluri,$^1$ K.C. Ku,$^2$ T. Wojtowicz,$^{3,4}$
  X. Liu,$^3$ J.K. Furdyna,$^3$ Y.B. Lyanda-Geller,$^5$
  N. Samarth,$^2$  and B. Nadgorny$^1$}   
\affiliation{$^1$Department of Physics and Astronomy, Wayne State
  University, Detroit, MI  48201\\
$^2$Department of Physics and Materials Research Institute,\\
The Pennsylvania State University, University Park, PA 16802\\
$^3$Department of Physics, University of Notre Dame, Notre Dame, IN  46556\\
$^4$Institute of Physics, Polish Academy of Sciences, 02-668 Warsaw, Poland\\
$^5$Department of Physics, Purdue University, West Lafayette, IN
  47907}

\date{\today}

\begin{abstract}
We investigate the applicability of spin polarization measurements
using Andreev reflection in a point contact geometry in heavily doped
dilute magnetic semiconductors, such as (Ga,Mn)As. While we observe
conventional Andreev reflection in non-magnetic (Ga,Be)As epilayers,
our measurements indicate that in ferromagnetic (Ga,Mn)As epilayers
with comparable hole concentration the conductance spectra can only be
adequately described by a broadened density of states and a reduced
superconducting gap. We suggest that these pair-breaking effects stem
from inelastic scattering in the metallic impurity band of (Ga,Mn)As
and can be explained by introducing a finite quasiparticle lifetime or
a higher effective temperature. For (Ga,Mn)As with 8\% Mn concentration
and 140~K Curie temperature we evaluate the spin polarization to be
$83\pm 17\%$.
\end{abstract}

\maketitle

{}The advance of semiconductor spintronics has revived a long-standing
interest in understanding the coupling of charge and spin in
semiconductors \cite{ArOr}. Ferromagnetic semiconductors \cite{Dietl},
\cite{Ohno/Samarth} are of central importance to semiconductor spintronics
since they have a conductivity compatible with that of conventional
semiconductors and the potential for a high intrinsic spin
polarization, thus providing promising conditions for efficient spin
injection into conventional semiconductors. In this context the
ferromagnetic semiconductor (Ga,Mn)As with Curie temperatures
routinely reproducible in the range of $140 \lesssim T_C \lesssim
170$K \cite{Ku}, \cite{Edmonds} stands out as
a well-studied model system \cite{Dietl}, \cite{Ohno/Samarth}. Furthermore,
(Ga,Mn)As has been successfully incorporated into a variety of spin
injection and spin transport devices \cite{GaMnAsjunction}. Although
measurements of the carrier (hole) spin polarization in this material
are immediately relevant to contemporary efforts in semiconductor
spintronics, systematic experiments probing this important quantity
are just beginning \cite{Braden}.

Spin polarization measurements can be carried out using the tunneling
geometry in superconductor (S)/insulator (I)/ferromagnet (FM)
structures \cite{Tedrow}; however, attempts to use this technique in
(Ga,Mn)As have thus far been unsuccessful \cite{Braden}. In spite of some
theoretical problems \cite{Kelly}, Andreev reflection (AR) in a FM/S
contact \cite{Soulen,Buhrman} provides a viable alternative to tunneling
for measurements of the spin polarization ($P$) in a variety of
materials, including ferromagnetic metals and metallic oxides
\cite{Zutic}. Recently, AR measurements in planar junctions of Ga/(Ga,Mn)As
have estimated a value of $ P\sim 85\%$ for samples with 5\% Mn and $T_C=65$~K
\cite{Braden}. Despite extensive attempts to make epitaxial S/(Ga,Mn)As
planar junctions with a variety of superconductors, useful data has
been obtained only in a limited number of Ga/(Ga,Mn)As samples
\cite{Braden}, suggesting extreme sensitivity to
the nature of the 
heterointerface.  Additionally, the planar geometry has unavoidable
limitations imposed by constraints on the materials growth, limiting
the post-growth modifications of the sample characteristics \cite{capping}
and thus restricting AR measurements to (Ga,Mn)As samples with
relatively low Curie temperatures ($T_C \sim 65$~K). Finally, the conductance
spectra in Ref. \cite{Braden} have been explained using a distribution of
the energy gaps in the Ga superconducting film. An alternative
interpretation of this data has also been suggested, involving a
distortion of the density of states in a superconductor \cite{Kant}. In
this context, the measurements of spin polarization in (Ga,Mn)As in
the conventional point contact Andreev reflection (PCAR) geometry are
vital for both extending the range of sample parameters as well as for
resolving different interpretation of the data.

In this Letter, we use PCAR to evaluate $P$ in (Ga,Mn)As epitaxial
layers with a high Curie temperature, $T_C=140$~K. In order to develop a
reliable interpretation of our study of (Ga,Mn)As, we first apply the
PCAR technique to a non-magnetic analogue of (Ga,Mn)As -- (Ga,Be)As --
with doping concentrations similar to those in the ferromagnetic
semiconductor ($ p \sim 10^{21} {\rm cm}^{-3}$). Our PCAR studies of
(Ga,Be)As yield 
the data that is well described by a {\it conventional} weak coupling
Blonder-Tinkham-Klapwijk (BTK) model \cite{BTK}. In contrast, the PCAR
experiments in ferromagnetic (Ga,Mn)As cannot be described by a simple
BTK model modified for the spin-polarized case \cite{Mazin}. The (Ga,Mn)As
data indicate a significant broadening of the density of states (DOS)
accompanied by a reduction of the bulk superconducting gap $\Delta_b$. We note
that these observations are not an intrinsic characteristic of
ferromagnetic semiconductors: for instance, PCAR measurements of
(In,Mn)Sb -- a higher mobility ferromagnetic semiconductor --
consistently yield the bulk superconducting gap with no DOS broadening
\cite{InMnSb}. This suggests that our observations in (Ga,Mn)As stem from
inelastic scattering in a low mobility ferromagnetic semiconductor.

Since the pioneering work of Kastalsky {\it et al.} \cite{Kastalsky}
most studies 
of AR in semiconductors have been carried out in a 2D geometry. This
is not surprising, as serious problems are anticipated for AR
experiments in a superconductor-semiconductor (S/Sm) junction in a 3D
geometry due to the high resistivity of semiconductors at low
temperatures and the presence of a Schottky barrier in most S/Sm
contacts. The Schottky barrier fundamentally limits the accuracy of
spin polarization measurements in ferromagnetic semiconductors by
strongly decreasing the probability of AR. To avoid both these
problems, we use heavily doped (Ga,Be)As and (Ga,Mn)As semiconductors
with metallic type conductivity and thus thin Schottky barriers, which
make highly transparent S/Sm junctions \cite{Schottky}.  

The models of
ferromagnetism in (Ga,Mn)As discussed in the literature invoke either
free valence hole \cite{Dietl,McDonald}, or impurity bands
\cite{DasSarma/Bhatt}. Both Mn and Be are nominally acceptors, with binding
energies of 113 and 28~meV, respectively. As a result, at low doping
concentrations transport in both (Ga,Be)As and (Ga,Mn)As at our
characteristic experimental temperature $T=1$~K must be described by
impurity bands. At high doping levels of $N \sim 10^{21} {\rm
  cm}^{-3}$ we assume 
that impurity disorder yields spacial modulation near the top of the
valence band, which results in a metallic state with ``ballistic''
propagation through the contact for (Ga,Be)As, and ``diffusive''
propagation for (Ga,Mn)As. Note that the energy scale determined by
kinetic and potential energies $e^2N^{1/3}/\varepsilon$ and
$\hbar^2N^{2/3}/2m$,
respectively, is of the order of 100~meV, where $\varepsilon$ is the dielectric
constant and $m$ is the effective mass. This is especially important for
(Be,Ga)As, which behaves as a conventional heavily doped semiconductor
in which the valence band is modulated by the impurity potential.

A number of 230~nm thick (Ga,Be)As samples with hole concentrations $p
= 8\times 10^{20} {\rm cm}^{-3}$ and $p = 5\times 10^{20} {\rm
  cm}^{-3}$ were grown by low-temperature (LT) 
molecular beam epitaxy (Riber 32 R\&D) on semi-insulating (001) GaAs
substrates. The 15~nm thick (Ga,Mn)As samples with a Mn composition of
8\% were grown in an EPI 930 system on n+, epi-ready (001) GaAs
substrates using conditions described elsewhere \cite{Ku}. Post-growth
annealing of the (Ga,Mn)As samples at 250 $^\circ$C yielded $T_C =140$~K and a
resistivity $\rho \sim 2 m\Omega \cdot {\rm cm}$ at 4.2~K \cite{Ku}. A
point contact  is established between the
sample and a mechanically polished Sn tip. Conductance ($dI/dV$) curves
were measured with the standard lock-in technique, as described in
detail in Ref. \cite{MnAs}, allowing us to monitor the characteristics of
same point contact from $\sim 1$~K to the critical temperature $T_c= 3.7$~K
of the Sn tip.

\begin{figure}
\includegraphics{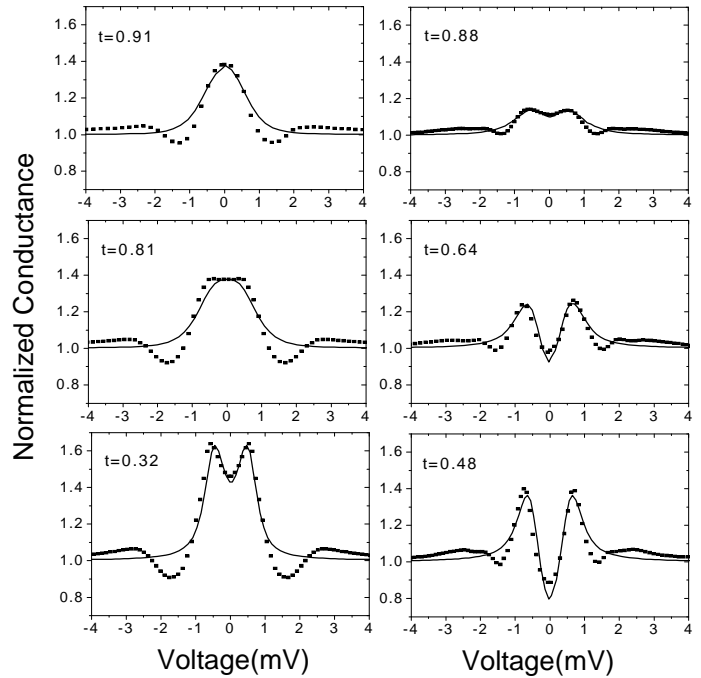}
\caption{\label{fig1}Two Sn/(Ga,Be)As contacts at different reduced
temperatures $t = T/T_c$ analyzed with the model of
Ref. \cite{Mazin}. Left panel: Contact resistance $R_c= 35\Omega$,
$Z \sim 0.45$; right panel: $R_c = 28\Omega$, $Z\sim 0.8$. A small dip
above $\Delta$ is due
to the proximity effect. }
\end{figure}

\begin{figure}
\includegraphics{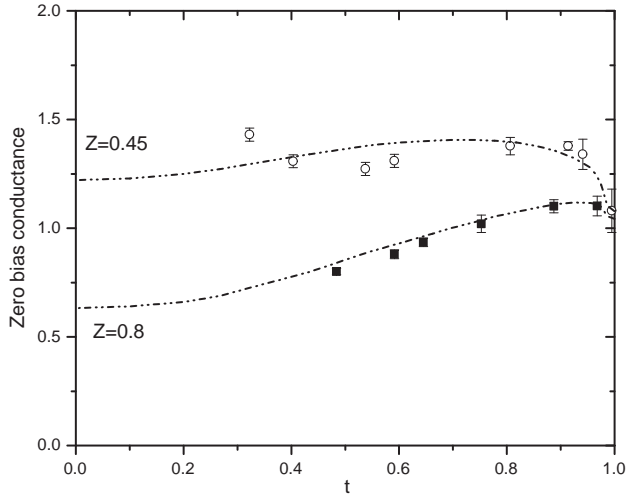}
\caption{\label{fig2}Zero bias conductance for the two contacts shown
  in Fig.~\ref{fig1}.  Dashed lines show the exact results obtained
  from the BTK  model for $Z = 0.45$ and $Z = 0.8$ respectively. }
\end{figure}

To study the properties of AR in non-magnetic semiconductors, we have
measured a series of temperature dependencies for a large number of
different Sn/(Ga,Be)As point contacts. In Fig.~\ref{fig1}, we show the
evolution of $dI/dV$ for two typical contacts in a sample with a hole
concentration $p = 8\times 10^{20} {\rm cm}^{-3}$ and a residual
resistivity $\rho \sim 150 \mu \Omega \cdot {\rm cm}$.  
Each
of the $dI/dV$ curves is analyzed independently using the model of
Ref. \cite{Mazin}, with the interface transparency of a contact
characterized by a dimensionless parameter $Z$. In this analysis, we use
the measured physical temperature of the contact and the corresponding
value of the Bardeen-Cooper-Schrieffer (BCS) gap $\Delta_b$. This procedure
results in a range of $Z$ for different contacts $0.4<Z<0.8$, with $Z$
practically temperature-independent for a given contact. The
resistivity and the measured carrier concentration for (Ga,Be)As yield
a mean free path $l\sim 10$~nm; this is comparable to the contact size
$d\sim 10$~nm
and suggests that the measurements occur in the ballistic transport
regime. Although complications may arise from the need to match wave
functions of different symmetry from (Ga,Be)As and Sn \cite{Kelly}, we will
attempt to describe the system phenomenologically following
Ref. \cite{BTK,ZuticSM}. For (Ga,Be)As we assume that the impurity and
the valence bands overlap, and thus we can still use light and heavy
holes to estimate the minimum $Z$-values, $Z=[(r-1)^2/4r]^{1/2}$,
which are due to the Fermi 
velocity mismatch $r$ between the superconductor (Sn) and (Ga,Be)As,
$r=v_{\rm Sn}/v_{\rm GaAs}$. Simple estimates of $r$ for light and
heavy holes, $r_l \sim m_{lh}(n_{\rm Sn}/n_{\rm GaAs})^{1/3}) \sim 1.7$
  and $r_h \sim m_{hh}(n_{\rm Sn}/n_{\rm GaAs})^{1/3}) \sim 4$, result
    in $Z_{hh} \sim 0.8$ and $Z_{lh} \sim 0.3$, 
in good agreement with the $Z$-values obtained from analyzing $dI/dV$
curves. The 
experimental zero-bias conductance for the two contacts shown in
Fig.~\ref{fig1}
and the two corresponding curves obtained {\it independently} from the BTK
model with $Z=0.45$ and $Z=0.8$ are shown in Fig.~\ref{fig2}. 
The surprisingly good agreement
between the data and the BTK model indicate the ``canonical'' AR,
typically observed in all-metal systems \cite{BTK}. Similar results were
obtained for (Ga,Be)As with $p= 5\times 10^{20}{ \rm cm}^{-3}$ and for
(In,Be)Sb 
\cite{InMnSb}. These results - in conjunction with the estimates of the
$Z$-values based on the Fermi velocity mismatch - suggest that the use of
highly doped semiconductors can minimize the role of the Schottky
barrier in these measurements. Simple estimates yield the Schottky
barrier thickness in this system of the order of several \AA, confirming
these conclusions. This is also consistent with the experimentally
observed symmetric and linear $I - V$ characteristics above
$\Delta_b$.  

\begin{figure}
\includegraphics{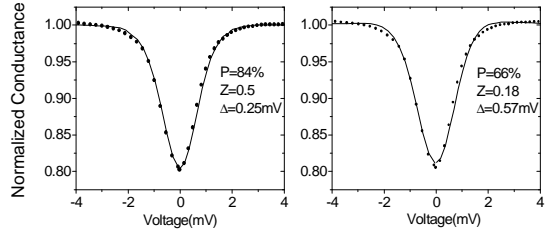}
\caption{\label{fig3}Fits to the modified BTK model for Sn/(Ga,Mn)As
  contact with  $R_c= 68 \Omega$ measured at $T =
  1.2$~K. $T^{\ast}=5.2$~K was used for both fits.  A reduced gap
  provides a better fit (left panel).  The variation of $\Delta$ results in
  different $P$. } 
\end{figure}

After
demonstrating that we can thoroughly understand the PCAR measurements
in the heavily-doped (Ga,Be)As, we now turn to the ferromagnetic
semiconductor (Ga,Mn)As with comparable carrier concentration. A
number of Sn contacts with (Ga,Mn)As epilayers have been
investigated. Qualitatively, all the contacts appear similar and
exhibit a zero-bias conductance that is significantly smaller than the
conductance at $V \gg \Delta/e$, suggesting a high spin polarization in
(Ga,Mn)As. However, analyzing the data is much more difficult compared
to both the non-magnetic case of (Ga,Be)As and the magnetic case of
(In,Mn)Sb \cite{InMnSb}. We find that AR in (Ga,Mn)As does not fit a model
of Ref. \cite{Mazin}, as all the experimental curves show a strong
broadening of the DOS and reduction of the superconducting gap (see
Fig.~\ref{fig3}).  

Our results on (Ga,Mn)As as well as (In,Mn)Sb \cite{InMnSb}
indicate that transport processes occurring in the semiconductor are
most likely responsible for the observed effects \cite{field}. In the
ballistic regime $l>d$, typical for most AR experiments, the conductance is
determined by the transparency of the FM/S interface with the bulk
superconducting gap $\Delta_b$. For (In,Mn)Sb \cite{InMnSb}, our
estimates show that 
holes are all in the ballistic regime. Thus the (In,Mn)Sb data is
easily interpreted in terms of a ballistic model \cite{Mazin}. In strong
contrast, transport in the (Ga,Mn)As impurity bands is ``diffusive'',
with $l \sim 2$~nm. While diffusive transport in AR experiments in
metals can be 
described by conventional theory \cite{Mazin}, in (Ga,Mn)As with $d
\sim 150$~nm the holes
spend significant time $t^\ast \sim d^2/D \sim 10^{-10}$~s within the
contact area, where $D$ is the
diffusivity. Hence the holes experience enhanced inelastic and
spin-flip scattering \cite{Alt-Ar,highspin}, with $t^\ast$
comparable to the 
hole scattering time $\tau_\varepsilon \sim 10^{-9} - 10^{-10}$~s due
to acoustic phonons and hole-hole
interaction. However, the observed broadened DOS and reduced
$\Delta_b$ both
require processes with shorter characteristic times $\sim 10^{-11}$~s, possibly
inelastic scattering off magnetic ions. We note that, while inelastic
scattering can explain the DOS broadening and gap reduction of the
superconductor, spin-flip scattering provides additional channels for
the Andreev current, introducing an uncertainty in spin polarization
measurements \cite{Falko}.  

We describe pair-breaking effects in the
Sn/(Ga,Mn)As contact using an empirical approach \cite{Dynes}, wherein we
account for inelastic scattering via an effective temperature, $T^\ast$, and a
reduced superconducting gap $\Delta$. Our approach provides a good description
of the experimental data, as seen in Fig.~\ref{fig3}. However, as we are unable
to evaluate $\Delta$ and $T^\ast$ from first principles, our model leads to the
uncertainty in  $\Delta$ and a broadened DOS. This in turn yields a
fairly large 
uncertainty in the extracted values of $P$ for (Ga,Mn)As, $P=83 \pm
17\%$. We have 
observed a qualitatively similar -- but quantitatively less
significant -- gap reduction and DOS broadening in ongoing PCAR
studies of the ferromagnetic semiconductor (Ga,Mn)Sb, whose mobility
is between that of (Ga,Mn)As and (In,Mn)Sb \cite{GaMnSb}. In that case, the
accuracy in determining the spin polarization is significantly better,
$P=57 \pm 5\%$ 
\cite{GaMnSb}. We note that, while we have a single critical temperature of
Sn, $T_c=3.7$~K, the observed spectra are qualitatively quite similar to
those in Ref. \cite{Braden}, suggesting that they may also be explained by
the gap reduction and DOS broadening. Our observations are consistent
with our conjecture that the PCAR measurements may suffer from
inelastic scattering effects that enhance the uncertainty in measuring
the spin polarization, particularly in highly spin-polarized materials
characterized by unconventional transport mechanism, such as
(Ga,Mn)As. This explanation is also in agreement with a recent
experiment in superconductor/normal metal nanostructures, in which Pt
impurities have been deliberately introduced at the
superconductor/normal metal interface to enhance inelastic scattering
\cite{privateBuhrman}.  

{\it Acknowledgments}: We thank R.A. Buhrman,
A.A. Golubov, E. Demler, V.I. Fal'ko, I.I. Mazin, S.A.Wolf, and
I. Zutic for useful discussions. The work is supported by the DARPA
SpinS through ONR grant N00014-02-1-0886 and NSF Career grant 0239058
(B.N.), by ONR grants N00014-99-1-0071, --0716, and N00014-99-1-1093
(N.S), and by the DARPA SpinS and NSF-NIRT Grant DMR02-01519 (J.K.F.).

\end{document}